\documentclass[mathpazo]{cicp}

\usepackage{amsmath}
\allowdisplaybreaks[4]
\usepackage{lipsum,soul}
\usepackage{bm}
\usepackage{graphicx}
\usepackage{graphics}
\usepackage{amsmath}
\usepackage{array}
\usepackage[caption=false]{subfig}
\usepackage{xcolor}
\usepackage{subfig}
\usepackage{hyperref}
\usepackage[capitalise]{cleveref}
\usepackage{textcomp}

\usepackage{epstopdf}
\usepackage{cases}
\usepackage{amssymb}
\usepackage{multirow}
\usepackage{siunitx}
\usepackage{cases}
\usepackage{tabularx}
\usepackage{nomencl}
\usepackage[top=1.2in, bottom=1.2in, left=1.1in, right=1.1in]{geometry}
\usepackage{tikz}
\usepackage{comment}
\usepackage{lineno,hyperref}
\modulolinenumbers[1]
\usepackage{booktabs}

\begin{document}

\title{A well-balanced lattice Boltzmann model for binary fluids based on the incompressible phase-field theory}

\author[Long Ju et.~al.]{Long Ju\affil{1},
       Peiyao Liu\affil{2}, Bicheng Yan\affil{4}, Jin Bao\affil{2}, Shuyu Sun\affil{1}\comma\corrauth and Zhaoli Guo\affil{3}\comma\corrauthB}
       
 \address{\affilnum{1}\ Computational Transport Phenomena Laboratory (CTPL), King Abdullah University of Science and Technology (KAUST), Thuwal, 23955-6900, Kingdom of Saudi Arabia. \\
 \affilnum{2}\ State Key Laboratory of Coal Combustion, Huazhong University of Science and Technology,Wuhan, 430074, China.\\
 \affilnum{3}\ Institute of Interdisciplinary Research for Mathematics and Applied Science, Huazhong University of Science and Technology, Wuhan 430074, China,\\
  \affilnum{4}\ Physical Science and Engineering Division, King Abdullah University of Science and Technology (KAUST), Thuwal 23955, Saudi Arabia.}
 \email{{\tt shuyu.sun@kaust.edu.sa} (S.~Sun)}
 \emails{{\tt zlguo@hust.edu.cn} (Z.~Guo)}

\date{\today}
\begin{abstract}
Spurious velocities arising from the imperfect offset of the undesired term at the discrete level are frequently observed in numerical simulations of equilibrium multiphase flow systems using the lattice Boltzmann equation (LBE) method. To capture the physical equilibrium state of two-phase fluid systems and eliminate spurious velocities, a well-balanced LBE model based on the incompressible phase-field  theory is developed. In this model, the equilibrium distribution function for the Cahn-Hilliard (CH) equation is designed by treating the convection term as a source to avoid the introduction of undesired terms, enabling achievement of possible discrete force balance. Furthermore, this approach allows for the attainment of a divergence-free velocity field, effectively mitigating the impact of artificial compression effects and enhancing numerical stability. Numerical tests, including a flat interface problem, a stationary droplet, and the coalescence of two droplets, demonstrate the well-balanced properties and improvements in the stability of the present model.
\end{abstract}

\maketitle
\section{Introduction}
Multiphase flows are frequently encountered in industrial operations and engineering applications. The description and prediction of the multiphase flow are difficult due to the complex interfacial behavior over a wide range of length and time scales~\cite{powell2008experimental,bozzini2003evaluation}. In order to comprehend them, numerous numerical simulation methods have been developed and have proven to be extremely effective~\cite{monaghan1995sph,koroteev2014direct,sakai2020recent,yang2016lattice,liang2014phase,yang2022free,zeng2022well,FENG2023111997,sun2020reservoir}. Among them, numerical schemes~\cite{li2013lattice,li2015lattice,zhang2019high,zhang2018discrete} based on the kinetic theory have been emerging as an appealing methodology in recent years, since they bridge the gap between macroscopic descriptions of multiphase dynamics and microscopic intermolecular interactions. In particular, the lattice Boltzmann equation (LBE) method has received particular attention for its concise and intuitive way of representing intermolecular interactions~\cite{li2016lattice,guo2013lattice}.
Although significant advances have been achieved in the development of multiphase LBE methods, some difficulties such as spurious velocity (SV) and numerical instability for systems with high density ratio and large viscosity ratio still exist~\cite{connington2012review}. SV is a phenomenon emerging in the vicinity of the phase interface. Theoretically, when a two-phase system is in equilibrium, the chemical potential of the system should remain constant, and the velocity field should be zero. However, the existing literature has indicated that the velocity near the interface cannot be completely eliminated, which is attributed to numerical reasons and is considered non-physical~\cite{wagner2003origin}. It is evident that the presence of SV could produce inaccurate density properties and may lead to unphysical phenomena in some situations, causing misunderstanding of the real physics.

Over the past few decades, significant efforts have been dedicated to identifying the underlying cause of SV and mitigating its impact~\cite{pooley2008eliminating,shan2006analysis}. Lee and Fisher~\cite{lee2006eliminating} found that the spurious velocities can be effectively minimized by utilizing the potential form of surface tension alongside the isotropic finite difference scheme in the context of the free energy LBE method. Cristea and Sofonea~\cite{cristea2003reduction} proposed that the use of the first-order upwind scheme for computing space derivatives in the evolution equation can lead to SV in the finite-difference LB equation. To address this issue, they introduced a correction force term designed to eliminate it. Subsequently, Guo et al.~\cite{guo2011force} concluded through rigorous mathematical analysis that in the free-energy based LB equation model the emergence of SV is a result of the imbalance between the surface tension force and the gradient of ideal gas pressure at a discrete level. Then Lou and Guo~\cite{lou2015interface} introduced a Lax-Wendroff-type LB equation to mitigate the impact of the above imbalance, which allows control of the magnitude of SV by modifying the Courant-Friedrichs-Lewy (CFL) number.  Although the above efforts are helpful for understanding the SV and reduce the magnitude, it is still difficult to completely eliminate SV. Most recently, Guo et al.~\cite{guo2021well} proposed a well-balanced (WB) LBE model, where the equilibrium distribution function is redefined such that an artificial pressure is included in its second-order moment, instead of the ideal gas pressure, and the modified force does not contain the gradient of ideal gas pressure which is present in the standard free-energy LBE model. As a result, the SV can be effectively eliminated to machine accuracy and the consistent interface profile and bulk densities can also be well captured in free-energy model. Based on this idea, Zheng et al.~\cite{zheng2021eliminating} developed a WB LBE model for incompressible two-phase flows based on the phase field theory, which successfully reduces SV to machine accuracy. It should be noted that the study conducted by Zheng et al.~\cite{zheng2021eliminating} was motivated by the work of Guo~\cite{guo2021well}, which involves the reconstruction of a novel LBE model for solving Navier-Stokes (NS) equations. However, the analysis (detailed in the following text) revealed that the presence of SV in the phase-field LBE model can be attributed to an imbalance between the excess term and the source term used to offset it on the discrete level in the LBE model used to solve the CH equation. Therefore, it is more reasonable to eliminate the SV by designing the LBE model of interface capture. Furthermore, it should be emphasized that the convection term in the macroscopic equation, $\bm{\nabla}\cdot(\bm{u}\phi)$, is in complete agreement with $\bm{u}\cdot\bm{\nabla}\phi$ when the fluid is incompressible, where $\bm{u}$ is the velocity of the fluid and $\phi$ donates the order parameters. However, due to the artificial compression effect of LBE method~\cite{kruger2009shear}, it is impossible to achieve a completely zero velocity divergence, thus leading to the fluctuations of order parameter in same phase, and in some cases the order parameter can be out of the physical range, which becomes particularly serious in scenarios involving large density ratios and viscosity ratio.
To overcome this difficulty, we propose a phase-field-based well-balanced LBE scheme for two-phase fluid flow. In the present model, the convection term with divergence-free velocity field is treated as a source term in the LBE, and the equation recovered from the LBE does not incorporate the supplementary hydrodynamic term. As a result, there is no requirement for an additional source term to balance it, thus preventing the occurrence of imbalance and enabling the reduction of spurious velocities to their accurate values. Furthermore, the reinstated divergence-free CH equation guarantees computational stability.

The remainder of the present paper is organized as follows. In~\cref{sec2}, the phase-field theory and governing equations for multiphase flows are presented, followed by the discussion of the origin of SV, then a WB-LBE model is proposed to eliminate SV. ~\cref{sec3} provides the numerical validation to test the performance of the proposed model. Finally, a summary is given in~\cref{sec:4}.

\section{Phase-field-based LB model for incompressible multiphase flows}~\label{sec2}
\subsection{Phase-field theory and governing equations}
In the phase-field theory, the equilibrium properties can be described by a Landau free energy functional, which can be expressed as a function of the order parameter $\phi$:~\cite{chen2008phase,jamet2002theory,zheng2015lattice}
\begin{equation}\label{energy_function}
\mathcal{F}(\phi)=\int \Psi(\phi, \nabla \phi) d\Omega_V=\int \left[\psi(\phi)+\frac{\kappa}{2}|\nabla \phi|^2\right]d\Omega_V,
\end{equation}
where $\Psi(\phi, \nabla \phi)$ represents the total free-energy density in the spatial region $\Omega_V$. $\psi(\phi)$ is the bulk free energy density, and $\kappa|\nabla \phi|$ accounts for the surface energy with $\kappa$ being a positive free-energy coefficient.
For simplicity but without loss of generality, the double-well form of bulk free-energy is used in the present work, which can be expressed as~\cite{guo2021well}:
\begin{equation}\label{bulk_energy}
\psi(\phi)=\beta(\phi-\phi_l)^2(\phi-\phi_v)^2,
\end{equation}
where $\beta$ is a constant parameter related to $\kappa$:
\begin{equation}\label{beta}
\beta=\frac{8\kappa}{W^2|\phi_l-\phi_v|^2},
\end{equation}
where $W$ is the interface thickness, $\phi_l$ and $\phi_v$ are the order parameters of liquid and vapor phases, respectively. It noted that the surface tension $\sigma$ can determined by the above parameters $\beta$ and $\kappa$ as:
\begin{equation}\label{suiface_tension}
\sigma=\frac{|\phi_l-\phi_v|^3}{6}\sqrt{2\kappa \beta}.
\end{equation}
The variation of free energy $\mathcal{F}$ with regard to order parameter is referred to as the chemical potential, i.e.,
\begin{equation}\label{chemical_potencial}
{\mu}\equiv\frac{\partial \mathcal{F}}{\partial \phi}=\frac{d \psi(\phi)}{d\phi}-\kappa\nabla^2\phi=4\beta(\phi-\phi_l)(\phi-\phi_v)(\phi-\frac{\phi_l+\phi_v}{2})-\kappa \nabla^2 \phi.
\end{equation}
The evolution of the order parameter can be governed by the CH equation,
\begin{equation}\label{CH_equation}
\frac{\partial \phi}{\partial t}+\bm{\nabla}\cdot (\phi \bm{u}) =\bm{\nabla}\cdot M_{\phi} (\bm{\nabla}\mu),
\end{equation}
where $M_{\phi}$ is the mobility coefficient, and $\bm{u}$ is the fluid velocity, which is determined by
the incompressible Navier-Stokes (NS) equations:
\begin{subequations}\label{6}
\begin{equation}
\bm{\nabla}\cdot \bm{u} =0,
\end{equation}
\begin{equation}
\partial_t(\rho\bm{u})+\bm{\nabla}\cdot(\rho \bm{uu})=-\bm{\nabla}p + \bm{\nabla}\cdot \left[\rho \nu (\bm{\nabla u}+\bm{\nabla u}^T)\right]+\bm{F_s},
\end{equation}
\end{subequations}
where $\rho$ is the fluid density, $p$ is the hydrodynamic pressure, $\nu$ is the kinematic viscosity and $F_s=-\phi\bm{\nabla} \mu$ donates the interaction force between different phases.
Obviously, the convection term in~\cref{CH_equation} can be simplified as $\bm{\nabla}\cdot (\phi\bm{u})=\bm{u}\cdot \bm{\nabla}\phi$ for incompressible flows, and the CH equation can can be written as,
\begin{equation}\label{CH_equation2}
\frac{\partial \phi}{\partial t}+\bm{u}\cdot\bm{\nabla}\phi=\bm{\nabla}\cdot M_{\phi} (\bm{\nabla}\mu).
\end{equation}

\subsection{Discussion for the source of SV}
In CH equation-based LBE models, the motion of the phase interface is driven by the gradient of the chemical potential.  SV usually comes from an imbalance of the chemical potential. Thus we prefer to start the analysis with the LB model for interface capturing, and here we take the incompressible multi-phase model proposed by Yang et al.~\cite{yang2016lattice} as an example, where the LBE for the CH equation can be written as
\begin{equation}\label{BGKyang}
f_i(\bm{x}+\bm{c}_i\delta_t, t+\delta_t)-f_i(\bm{x},t)=-\frac{1}{\tau_f}\left[f_i(\bm{x},t)-f_i^{eq}(\bm{x},t)\right]+\delta_t
\left[1-\frac{1}{2\tau_f}\right]F'_i,
\end{equation}
where $f(\bm{x}, t)$ donates the particle distribution function, referring to a cluster of particles located at position $\bm{x}$ at time t. $\tau_f$ is the relaxation time. $f_i^{eq}(\bm{x},t)$ is the equilibrium distribution function. For the two-dimensional-nine-velocity (D2Q9) model, $f_i^{eq}$ is given by
\begin{equation}\label{H_iyang}
f_i^{eq}=
\begin{cases}
    \phi-(1-\omega_0)\alpha \mu+\omega_i \phi s_i(\bm{u}),& i=0,\\
    \omega_i\alpha \mu+\omega_i\phi s_i(\bm{u}),& i\neq 0,
\end{cases}
\end{equation}
with
\begin{equation}
    s_i(\bm{u})=\frac{\bm{c}_i\cdot \bm{u}}{c_s^2}+\frac{\bm{uu}:(\bm{c}_i\bm{c}_i-c_s^2I)}{2c_s^4}.
\end{equation}
where $\omega_i$ is the weight coefficient, which can be given by $\omega_0=4/9$, $\omega_{1-4}=1/9$ and $\omega_{5-8}=1/36$. $\alpha$ is an adjustable parameter related to the mobility $M_{\phi}$. and the discrete velocities $\bm{c}_i$ are
\begin{equation}
\bm{c}_i=
\begin{cases}
(0,0)c,& i=0\\
(\text{cos}[(i-2)\pi/2],\text{sin}[(i-2)\pi/2])c,& i=1\sim4,\\
(\text{cos}[(i-5)\pi/2+\pi/4],\text{sin}[(i-5)\pi/2+\pi/4])c,& i=5\sim8.\\
\end{cases}
\label{eq:ci}
\end{equation}
where $c=\delta_x/\delta_t$ is the lattice speed with $\delta_x$ and $\delta_t$ representing the grid spacing and the time increment, respectively, and $c_s^2=c/\sqrt{3}$ is the lattice sound speed.
The zeroth- through second-order moments of $f^{eq}_i$ are
\begin{align}\label{sumfeq}
    \sum_if_i^{eq}=\phi, \quad \sum_i\bm{c}_if_i^{eq}=\phi \bm{u}, \quad \sum_i\bm{c_i}\bm{c_i}f_i^{eq}=\phi\bm{uu}+c_s^2\alpha \mu.
\end{align}
By applying the Chapman-Enskog (CE) expansion~\cite{guo2013lattice} to the above LBE, we can obtain the following macroscopic equation,
\begin{equation}\label{eqlambda}
    \partial_t \phi+\bm{\nabla}\cdot (\phi \bm{u})=M_{\phi}\bm{\nabla}\cdot (\bm{\nabla}\mu+\Lambda_1-\Lambda_2),
\end{equation}
with
\begin{subequations}\label{cheqlam}
    \begin{equation}\label{lambda1}
        \Lambda_1=\partial_{t0}(\phi \bm{u})+\bm{\nabla}\cdot (\phi \bm{uu})=\frac{\phi}{\rho}(\bm{F}_s-\bm{\nabla}p),
    \end{equation}
    \begin{equation}\label{lambda2}
        \Lambda_2=\sum_i\bm{c}_iF'^{(0)}_i=\frac{\phi}{\rho}(\bm{F}_s-\bm{\nabla}p).
    \end{equation}
\end{subequations}
Although mathematically the terms $\Lambda_1$ and $\Lambda_2$ are equal and they could cancel each other out, their discrete version my not be, because $\Lambda_1$ is induced by the collision-streaming process of LBE, while $\Lambda_2$ is artificial.  The imbalance of these two terms on discrete level can lead to the contamination of the chemical potential, which further affects the hydrodynamic balance and finally leads to the SV.

\subsection{A well-balanced phase-field LBE model}
Based on the previous analysis, the equilibrium distribution function is usually designed with velocity-related term $s_i(\bm{u})$ to recover the convection term $\bm{\nabla} \cdot (\phi \bm{u})$. As a defect, an undesired term $\Lambda_1$ is introduced, thus the additional term $\Lambda_2$ should be used to cancel it. The different truncation errors in these two terms are the main reason for the SV.
Therefore, if we remove the velocity-related term $s_i(\bm{u})$, the term $\Lambda_1$ does not appear, and the artificial one $\Lambda_2$ is not yet required, and the imbalance can be avoided. The convection term is then regarded as a separate source term, which is inspired by the work of Huang et al.~\cite{huang2015lattice}
Following this idea, we will present a simple, accurate, and robust two-phase well balanced model in this subsection. Emphasizing simplicity and computational efficiency, our model is constructed using the single-relaxation-time, also known as the BGK method~\cite{zou1997pressure}, which can be easily extended to the advanced multiple-relaxation-time version.

\subsubsection{LBE for interface capturing}
The LBE for the CH equation can be expressed as~\cite{bin2005new},
\begin{equation}\label{BGK}
f_i(\bm{x}+\bm{c}_i\delta_t, t+\delta_t)-f_i(\bm{x},t)=-\frac{1}{\tau_f}\left[f_i(\bm{x},t)-f_i^{eq}(\bm{x},t)\right]+\delta_t {F}_i(\bm{x},t)+\frac{1}{2}\delta_t^2\partial_t F_i(\bm{x},t),
\end{equation}
with the equilibrium distribution function $f_i^{eq}$ being defined as
\begin{equation}\label{H_i}
f_i^{eq}=
\begin{cases}
    \phi-(1-\omega_0)\alpha \mu,& i=0,\\
    \omega_i\alpha \mu,& i\neq 0,
\end{cases}
\end{equation}
To recover the CH equation exactly, the source term $F_i$ is designed as~\cite{guo2002discrete}
\begin{equation}\label{F_i}
F_i(\bm{x},t)=\omega_i(\bm{u}\cdot \bm{\nabla} \phi)\left[-1+\frac{\bm{I}:(\bm{c}_i\bm{c}_i-c_s^2\bm{I})}{2c_s^2}\right],
\end{equation}
The order parameter is computed by
\begin{equation}\label{order-parameter}
\phi=\sum_if_i.
\end{equation}
It is noted that $f_i^{eq}$ and $F_i$ satisfy the following moments:
\begin{subequations}\label{sumfeqandF}
\begin{equation}\label{sumfeqandF1}
    \sum_if_i^{eq}=\phi, \quad \sum_i\bm{c}_if_i^{eq}=0, \quad \sum_i\bm{c_i}\bm{c_i}f_i^{eq}=c_s^2\alpha \mu,
\end{equation}
\begin{equation}\label{sumfeqandF2}
    \sum_iF_i=-\bm{u}\cdot \bm{\nabla}\phi, \quad \sum_i\bm{c}_iF_i=0, \quad \sum_i\bm{c_i}\bm{c_i}F_i=0.
\end{equation}
\end{subequations}
The above LBE model can be analyzed by applying the CE expansion~\cite{guo2013lattice}, and the multiscale expansions are given as
\begin{subequations}\label{expsion}
    \begin{equation}
        \bm{f}_i=\bm{f}_i^{(0)}+\epsilon \bm{f}_i^{(1)}+\epsilon^2 \bm{f}_i^{(2)}+...,
    \end{equation}
    \begin{equation}
        \partial_t=\epsilon\partial_{t0}+\epsilon^2\partial_{t1},\quad \bm{\nabla}=\epsilon \bm{\nabla_0},\quad F'_i=\epsilon F'^{(0)}_i,
    \end{equation}
\end{subequations}
Using the Taylor expansion in~\cref{BGK}, one can obtain
\begin{equation}\label{BGKT}
D_if_i+\frac{\delta_t}{2}D_i^2f_i=-\frac{1}{\delta_t\tau_f}(f_i-f_i^{eq})+F_i+\frac{\delta_t}{2}\partial_t F_i.
\end{equation}
Submitting~\cref{expsion} into~\cref{BGKT}, we can obtain that
\begin{subequations}\label{expsionsigmapresent}
    \begin{equation}\label{exp1pre}
        O(\epsilon^0): \quad f_i^{(0)}=f_i^{eq},
    \end{equation}
    \begin{equation}\label{exp2pre}
        O(\epsilon^1): \quad D_{0i}f_i^{(0)}=-\frac{1}{\delta_t \tau_f}f_i^{(1)}+F_i^{(0)},
    \end{equation}
    \begin{equation}\label{exp3pre}
        O(\epsilon^2): \quad \partial_{t1}f_i^{(0)}+D_{0i}f_i^{(1)}+\frac{\delta_t}{2}D_{0i}^{2}f_i^{(0)}=-\frac{1}{\delta_t \tau_f}f_i^{(2)}+\frac{\delta_t}{2}\partial_{t0}F_i^{(0)}.
    \end{equation}
\end{subequations}
Submitting~\cref{exp2pre} into~\cref{exp3pre}, one can obtain
\begin{equation}\label{exp23pre}
    \partial_{t1}f_i^{(0)}+\tau_f \delta_t D_{0i}F^{(0)}_i+\left(\frac{\delta_t}{2}-\delta_t \tau_f\right)D_{0i}^2 f_i^{(0)}=-\frac{1}{\delta_t \tau_f}f_i^{(2)}+\frac{\delta_t}{2}\partial_{t0} F_i^{(0)}.
\end{equation}
Combining~\cref{sumfeqandF}, and taking the zeroth-order moment of~\cref{exp2pre} and~\ref{exp23pre}, we can obtain
\begin{subequations}\label{recovereqpre}
    \begin{equation}\label{req1pre}
        \partial_{t0}\phi=-\bm{u}\cdot \bm{\nabla}_0\phi,
    \end{equation}
    \begin{equation}\label{req2pre}
        \begin{split}
             \partial_{t1}\phi+\left(\frac{\delta_t}{2}-\delta_t \tau_f\right) \bigg[\partial_{t0}\left(\partial_{t0}\phi+\bm{u}\cdot \bm{\nabla}_0 \phi\right)+\bm{\nabla}_0\cdot\bm{\nabla}_0(c_s^2\alpha\mu)\bigg]=0.
        \end{split}
    \end{equation}
\end{subequations}
Then the CH equation can be recovered as
\begin{equation}\label{CH_equation3}
\frac{\partial \phi}{\partial t}=\bm{\nabla}\cdot M_{\phi} (\bm{\nabla}\mu)-\bm{u}\cdot\bm{\nabla}\phi,
\end{equation}
with the mobility given by
\begin{equation}\label{mobility}
M_{\phi}=c_s^2\alpha(\tau_f-0.5)\delta_t.
\end{equation}

Since the final recovered governing equation is consistent with the CH equation without any addition term, it is expected that it will be able to preserve the well-balanced property.
In addition, it should be noted that the divergence-free condition can be achieved naturally using the above LBE model. Thus, the fluctuation of the order parameter caused by artificial compression effects can be filtered out, which could improve the numerical stability. We will demonstrate this point in numerical simulations.

\subsubsection{LBE for hydrodynamics}
The LBE for the incompressible Navier-Stokes equations, can be written as:
\begin{equation}\label{BGK-NS}
g_i(\bm{x}+\bm{c}_i\delta_t, t+\delta_t)-g_i(\bm{x},t)=-\frac{1}{\tau_g}\left[g_i(\bm{x},t)-g_i^{eq}(\bm{x},t)\right]+\delta_t {G}_i(\bm{x},t),
\end{equation}
where $g_i(\bm{x},t)$ represents the distribution function for solving the flow field, $g_i^{eq}(\bm{x},t)$ is the  equilibrium distribution function, which can be expressed as
\begin{equation}\label{G_i}
g_i^{eq}=
\begin{cases}
    \frac{p}{c_s^2}(\omega_i-1)+\rho {s}_i(\bm{u}),& i=0,\\
    \frac{p}{c_s^2}\omega_i+\rho {s}_i(\bm{u}),& i\neq 0,
\end{cases}
\end{equation}
with ${s}_i(\bm{u})$ being defined as
\begin{equation}\label{Siu}
{s}_i(\bm{u})=\omega_i\left[\frac{\bm{c}_i\cdot \bm{u}}{c_s^2}+\frac{(\bm{c}_i\cdot \bm{u})^2}{2c_s^4}-\frac{\bm{u}\cdot\bm{u}}{2c_s^2}\right],
\end{equation}
$G_i(\bm{x},t)$ in ~\cref{BGK-NS} symbolizes the force distribution function, which is given by~\cite{liang2018phase}
\begin{equation}\label{GGi}
G_i=(1-\frac{1}{2\tau_g})\omega_i\left[\bm{u}\cdot \bm{\nabla}\rho+\frac{\bm{c}_i\cdot \bm{F}}{c_s^2}+\frac{(\bm{u\nabla}\rho:(\bm{c}_i\bm{c}_i-c_s^2\bm{I})}{c_s^2}\right],
\end{equation}
where $\tau_g$ is the dimensionless relaxation time related to the viscosity, and $\bm{F}=\bm{F}_s+\bm{G}$ represents the total force with $\bm{G}$ being the body force. The macroscopic quantities $\bm{u}$ and $p$ can be evaluated as
\begin{subequations}
    \begin{equation}
        \rho\bm{u}=\sum_i\bm{c}_ig_i+0.5\delta_t\bm{F},
    \end{equation}
    \begin{equation}
        p=\frac{c_s^2}{(1-\omega_0)}\left[\sum_{i\neq 0}g_i+0.5\delta_t\bm{u}\cdot \bm{\nabla}\rho+\rho s_0(u)\right],
    \end{equation}
\end{subequations}
Based on the Chapman-Enskog analysis, the NS equations can be recovered from~\cref{BGK-NS} with the fluid kinematic viscosity determining by
\begin{equation}
    \nu=c_s^2(\tau_g-0.5)\delta_t.
\end{equation}

In a two-phase system, the viscosity is no longer homogeneous as it exhibits a discontinuity at the liquid-gas interface. Various approaches exist to handle the viscosity across this interface, including linear function of the order parameter~\cite{liang2018phase}, inverse linear form~\cite{zu2013phase} and exponential form~\cite{langaas2000lattice}. And in this work, a simple linear form as used for density is adopted:
\begin{equation}\label{vis}
\nu=\phi (\nu_l-\nu_v)+\nu_g.
\end{equation}

To ensure accurate numerical iterations, it is necessary to discretize the gradient and Laplace operators in the model using appropriate difference schemes. In this case, the gradient term is computed using the second-order isotropic central scheme for the sake of simplicity,
\begin{equation}\label{gradient}
\bm{\nabla}\Psi(\bm{x})=\sum_{i\neq 0}\frac{\omega_i\bm{c}_i\Psi(\bm{x}+\bm{c}_i\delta_t)}{c_s^2\delta_t},
\end{equation}
and the Laplace operator is calculated by
\begin{equation}\label{Laplace}
{\nabla}^2\Psi(\bm{x})=\sum_{i\neq 0}\frac{2\omega_i[\Psi(\bm{x}+\bm{c}_i\delta_t)-\Psi(\bm{x})]}{c_s^2\delta^2_t},
\end{equation}
where $\Psi$ is any physical variable~\cite{li2013lattice}.

\section{Numerical tests and discussion}~\label{sec3}
This section aims to validate the accuracy and robustness of the present WB-LB model. Detailed comparisons between the present results and analytical solutions or some available results will be presented to demonstrate the reliability of the present model.

\subsection{Flat interface}

The first test case is a flat interface problem, in which a computation domain of size $L_x\times L_y=32\times128$ is utilized. The liquid slab is placed in the central region of ($0.25L_y<y<0.75L_y$), while the remaining area is occupied by the vapor phase. To ensure continuity, periodic boundary conditions are implemented at all four boundaries. The order parameter at equilibrium in this plane interface problem exhibits a hyperbolic tangent profile, which can be expressed as~\cite{guo2021well}
\begin{equation}\label{flatinitial}
\phi(y)=\phi_v+0.5(\phi_l-\phi_v)\left[\tanh\left(\frac{2z_1}{W}\right)-\tanh\left(\frac{2z_2}{W}\right)\right],
\end{equation}
where $z_1=y-y_1$ and $z_2=y-y_2$ are the distances to the two interfaces with $y_1=0.25L_y$ and $y_2=0.75L_y$. The above initial smooth equilibrium profile is adopted in all testing cases.
In the simulations, the density and the kinematic viscosity of vapor phase chosen as $\rho_v=1$ and $\nu_v=0.1$, and cases with different density ratio and viscosity ratio are tested. The mobility of the order parameter $M_{\phi}$ is set as $M_{\phi}=0.1$ and the adjustable parameter $\alpha=1$. Surface tension $\sigma=0.005$, while $\beta$ and $\kappa$ can be determined by~\cref{beta} and ~\cref{suiface_tension} with interface thickness set as $W=4$.

\begin{figure}[ht]
     \centering
    \subfloat[]{\includegraphics[width=0.42\textwidth]{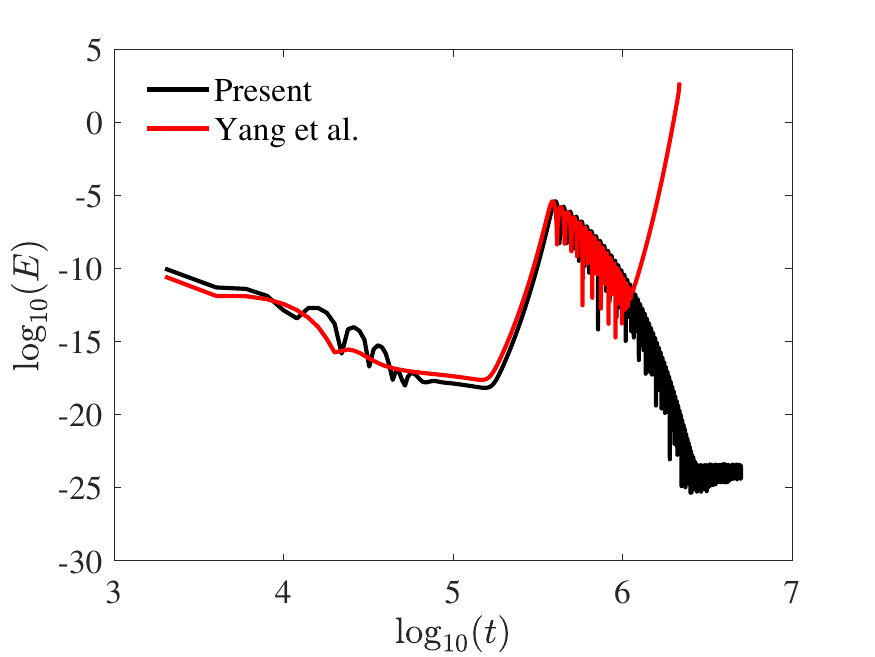}}~~
    \subfloat[]{\includegraphics[width=0.42\textwidth]{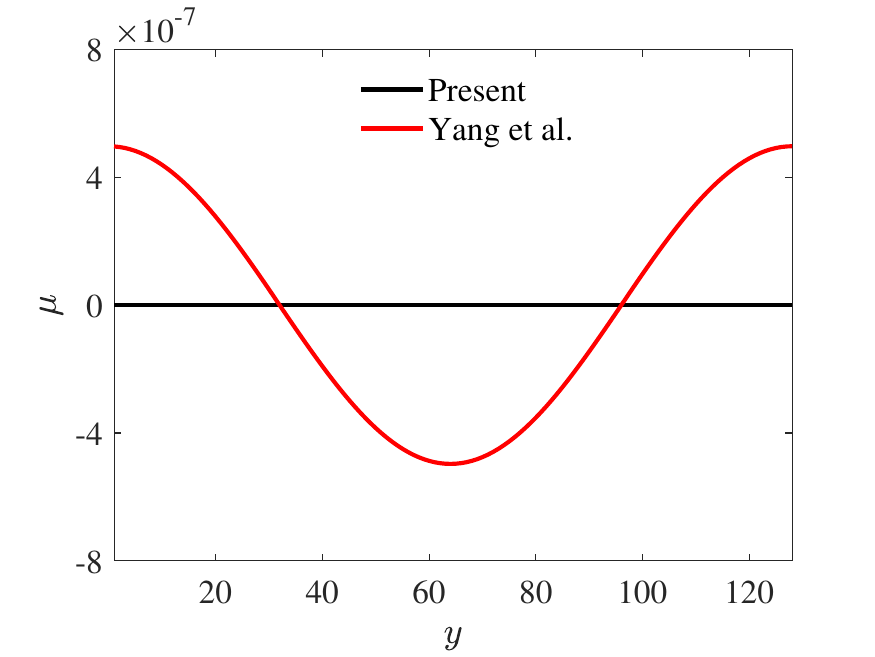}}~~\\
    \subfloat[]{\includegraphics[width=0.42\textwidth]{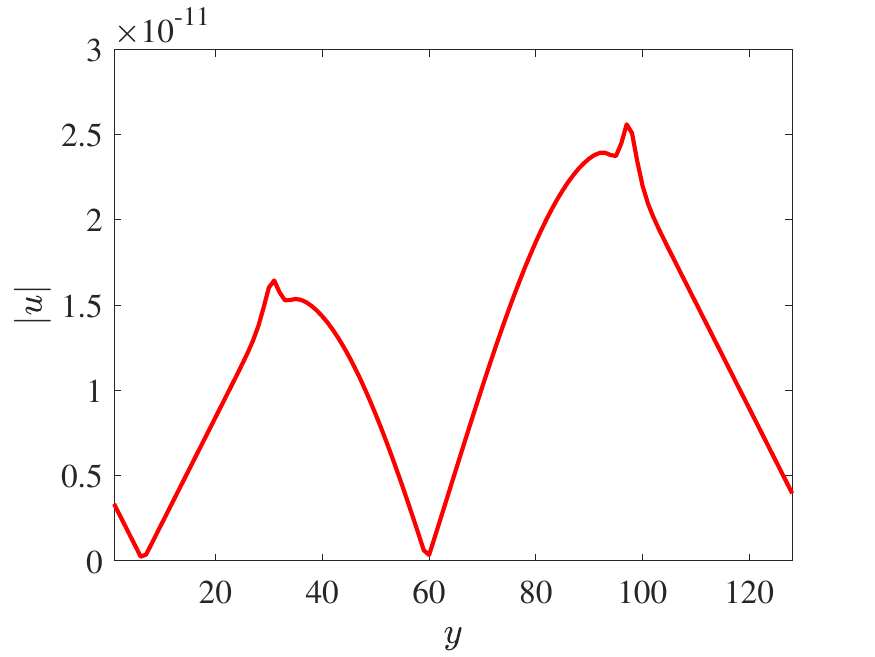}}~~
    \subfloat[]{\includegraphics[width=0.42\textwidth]{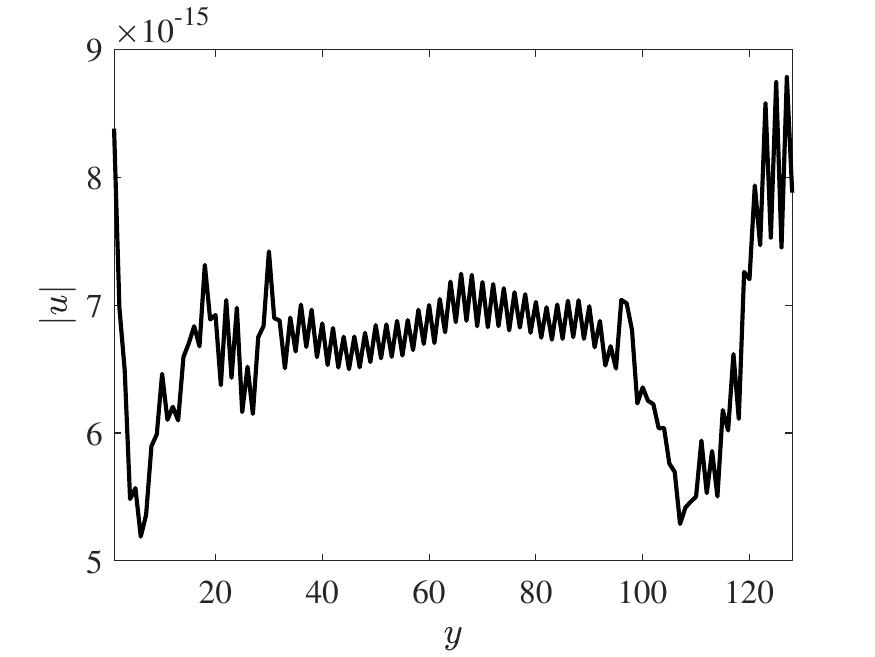}}~~
     \caption{Comparison results of the flat interface problem between the present model and Yang's model, $\nu_l/\nu_v=1$ and $\rho_l/\rho_v=10$. (a) The time history of the total kinetic energy, (b) chemical potential along $x=0.5L_x$ with the lowest total kinetic energy, (c) velocity profile along $x=0.5L_x$ obtained by the LB model of Yang et al.~\cite{yang2016lattice} with the lowest total kinetic energy, (d) steady velocity profile along $x=0.5L_x$ obtained by the present model.}
     \label{flatcomparion}
\end{figure}
A comparison with $\rho_l/\rho_v=10$ and $\nu_l/\nu_v=1$ is firstly conducted. ~\cref{flatcomparion} shows comparison results between the present model and Yang et al.'s model~\cite{yang2016lattice}. As illustrated in ~\cref{flatcomparion}(a), both the time history of $E=\frac{1}{2}\int \rho|\bm{u}|^2 d\bm{x}$ obtained from two models exhibit decrease with time initially, and then gradually increase near the magnitude of $O(10^{-5})$. For the results from Yang et al.'s model~\cite{yang2016lattice}, the total kinetic energy $E$ decreases first and then increases rapidly, after reaching the highest point, which indicates that the steady state cannot be reached in this case. While it is clearly that the total kinetic energy of the present model drops all the way down from the highest point and eventually reaches the order of $O(10^{-25})$.
The chemical potential distributions are shown in ~\cref{flatcomparion}(b), it is obvious that the chemical potential predicted by the present model is nearly constant, while that obtained by the reference model varies at the order of $10^{-7}$.
~\cref{flatcomparion}(c) shows the velocity profile at the lowest total kinetic energy obtained from Yang et al.'s model~\cite{yang2016lattice}, and ~\cref{flatcomparion}(d) gives the steady velocity profile along $x=0.5L_x$ obtained by the present model. It can be seen that the SV of the present model has been eliminated to machine accuracy.

\begin{figure}[ht]
     \centering
    \subfloat[]{\includegraphics[width=0.3\textwidth]{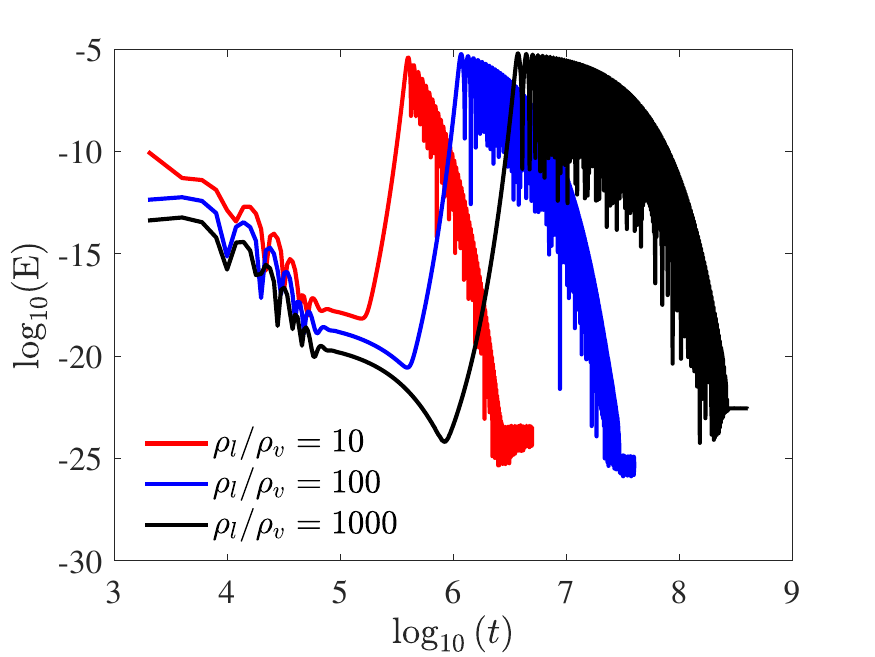}}~~
    \subfloat[]{\includegraphics[width=0.3\textwidth]{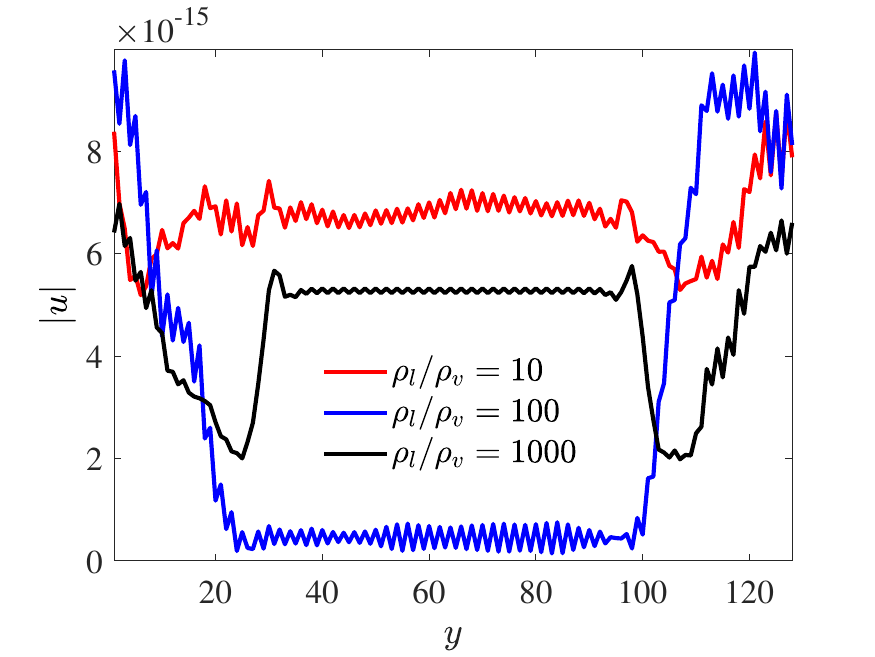}}~~
    \subfloat[]{\includegraphics[width=0.3\textwidth]{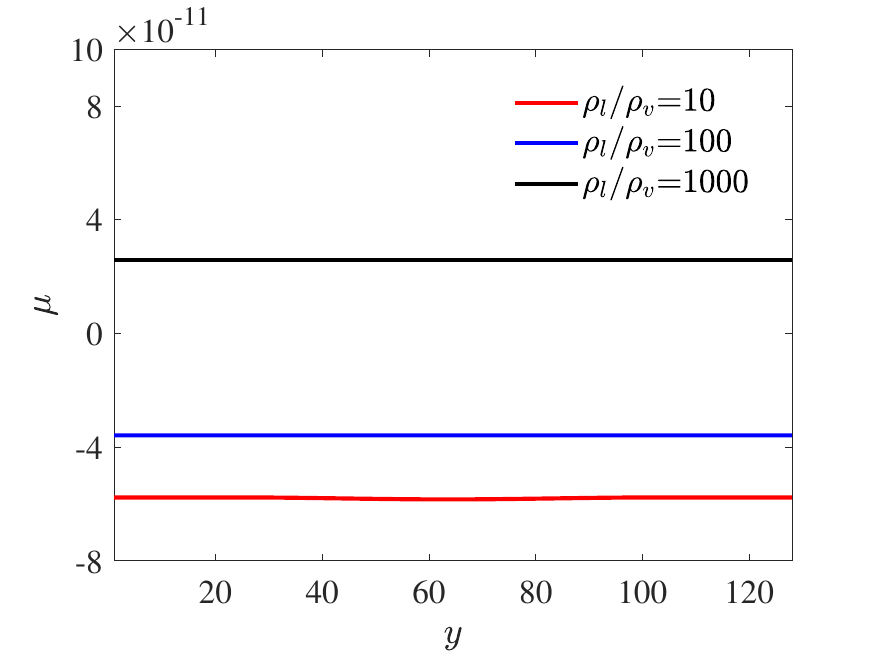}}~~
     \caption{Results of the flat interface problem with $\nu_l/\nu_v=1$ and $\rho_l/\rho_v=10,100,1000$,(a) The time history of the total kinetic energy, (b) velocity profile and (c) chemical potential at steady state.}
     \label{flatrho}
\end{figure}
Some other cases are further carried out to test the performance of the present model. ~\cref{flatrho}(a) plots the time history of the total kinetic energy for the cases of $\nu_l/\nu_v=1$ and three different density ratios ($\rho_l/\rho_v=10,100,$ and 100). As shown, in all cases with different density ratio, the kinetic energy are first reduced, and then gradually increase near the magnitude of $O(10^{-5})$, but finally they all reduce to about $O(10^{-25})$. The velocity profiles and the chemical potential distributions are shown in~\cref{flatrho}(b) and (c), it can be seen that the SV at steady state is on the order of $10^{-15}$, and the chemical potential is nearly constant, which imply that the well-balanced property of the present model.

\begin{figure}[ht]
     \centering
     \subfloat[]{\includegraphics[width=0.3\textwidth]{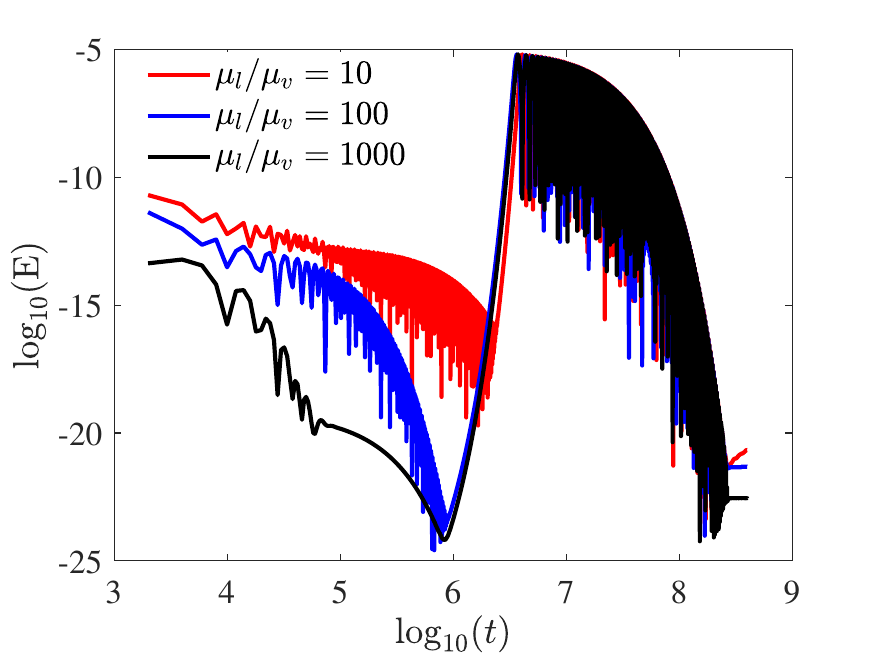}}~~
     \subfloat[]{\includegraphics[width=0.3\textwidth]{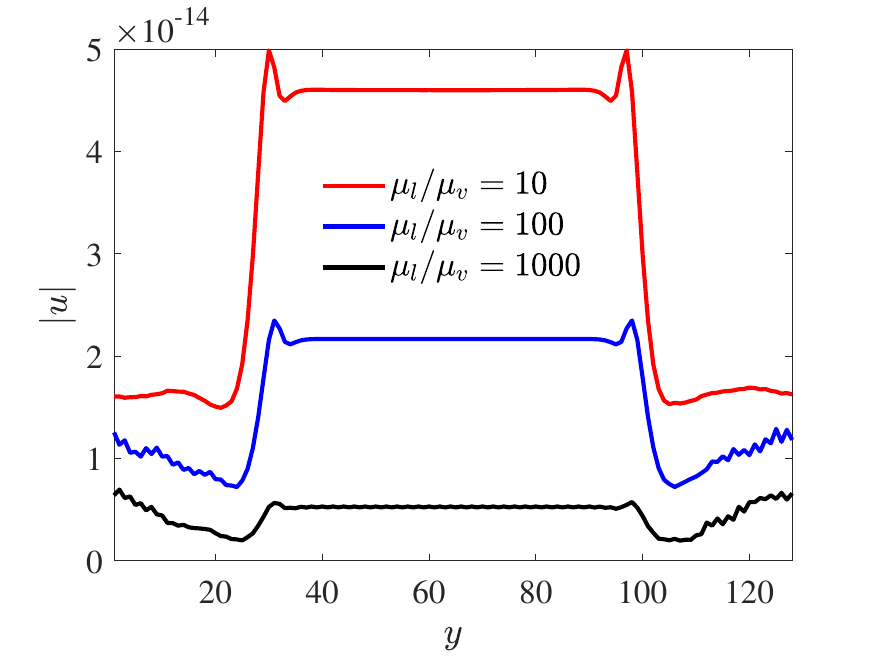}}~~
     \subfloat[]{\includegraphics[width=0.3\textwidth]{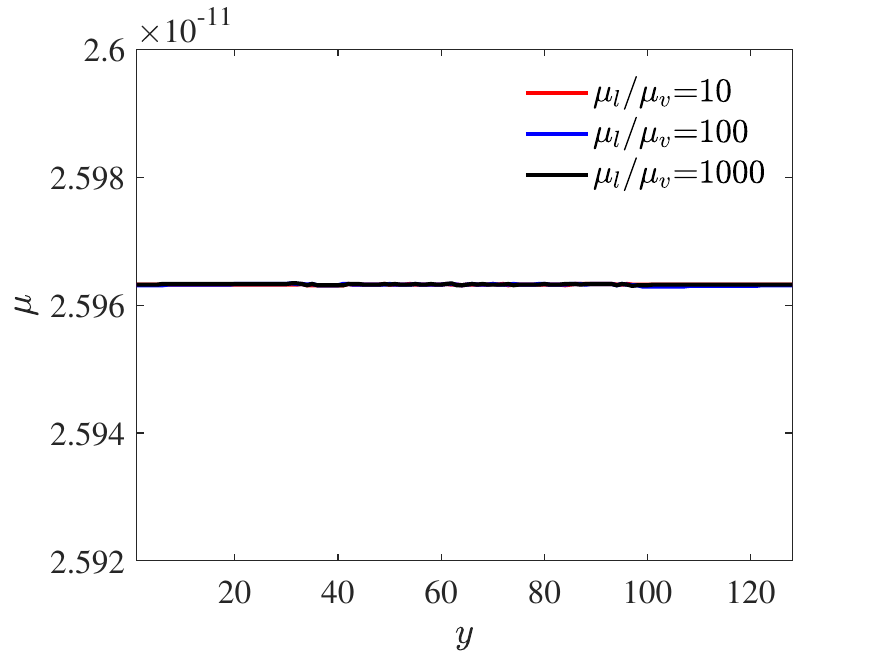}}~~
     \caption{Results of the flat interface problem with $\rho_l/\rho_v=1000$ and $\mu_l/\mu_v=10,100,1000$,(a) The time history of the total kinetic energy, (b) velocity profile and (c) chemical potential at steady state.}
     \label{flatmu}
\end{figure}

Several cases with different viscosity ratios at a fixed density ration $\rho_l/\rho_v=1000$ are then tested to prove the well-balanced property of the present model. Numerical results are shown in~\cref{flatmu}.
It can be seen that the kinetic energy $E$ are finally reduced to the order of $O(10^{-23})$ and the magnitude of SV at steady state is on the order of $10^{-14}$, and the chemical potential is also nearly constant.

\subsection{Stationary droplet}
We now employ the present LBE model to simulate the two-dimensional stationary circular droplet with radius $R$ in a square domain of size $L\times L=128\times 128$. Initially, the droplet is placed at the center area, and the rest of the domain is filled with its vapor. Periodic boundary conditions are applied to all surrounding boundaries, and the initial density profile is set as~\cite{yang2022free,zeng2022well}
\begin{equation}\label{case1}
\rho(x,y)=\frac{\rho_l+\rho_v}{2}+\frac{\rho_l-\rho_v}{2}\tanh\left(\frac{2\left(R-\sqrt{(x-x_c)^2+(y-y_c)^2}\right)}{W}\right),
\end{equation}
where $(x_c, y_c)=(0.5L, 0.5L)$ is the center position of the square domain. The radius of the droplet is set to $R=0.25L$. The density and kinematic viscosity of the vapor phase are $\rho_v=1$ and $\nu_v=0.1$, respectively.  Other parameters such as the mobility of the order parameter $M_{\phi}$, the adjustable parameter $\alpha$, the surface tension $\sigma$, and the thickness of the interface $W$ are set the same as in the flat interface problem.

\begin{figure}[ht]
     \centering
    \subfloat[]{\includegraphics[width=0.42\textwidth]{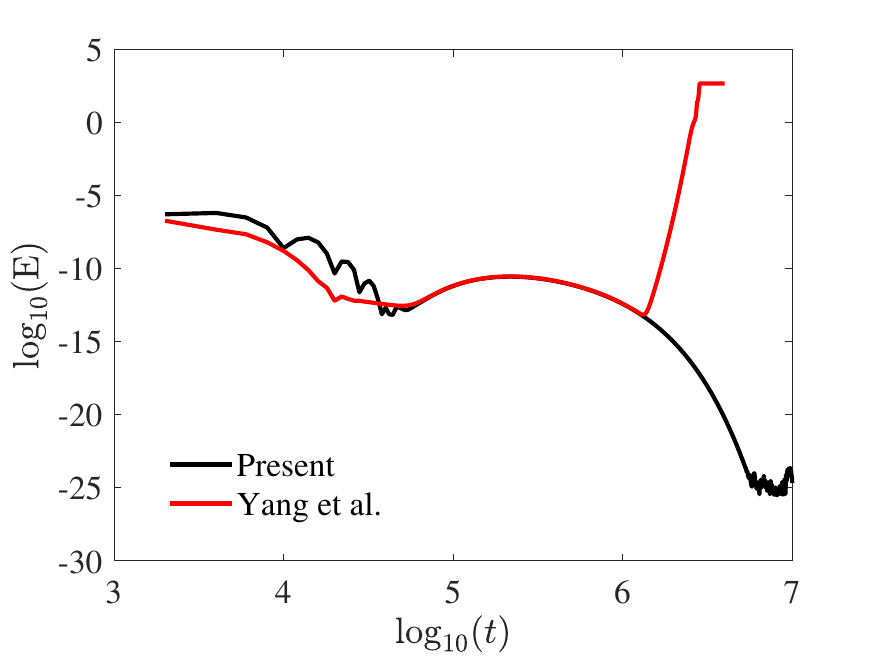}}~~
    \subfloat[]{\includegraphics[width=0.42\textwidth]{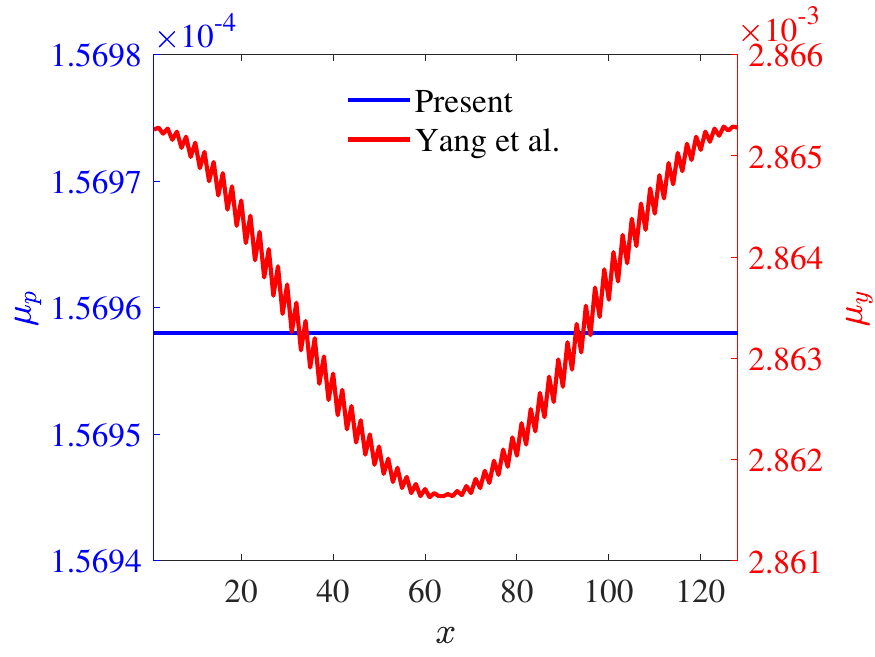}}~~\\
    \subfloat[]{\includegraphics[width=0.42\textwidth]{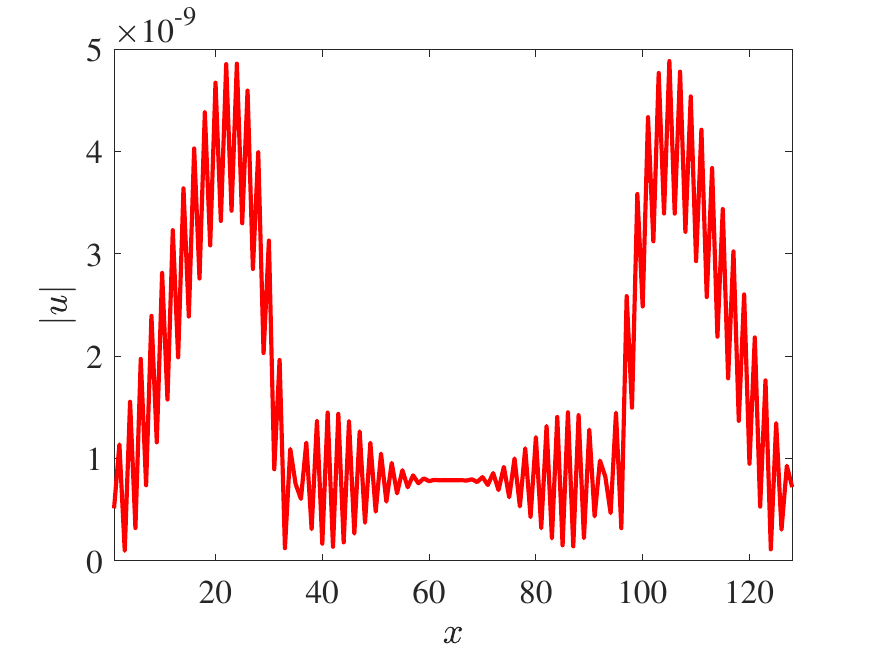}}~~
    \subfloat[]{\includegraphics[width=0.42\textwidth]{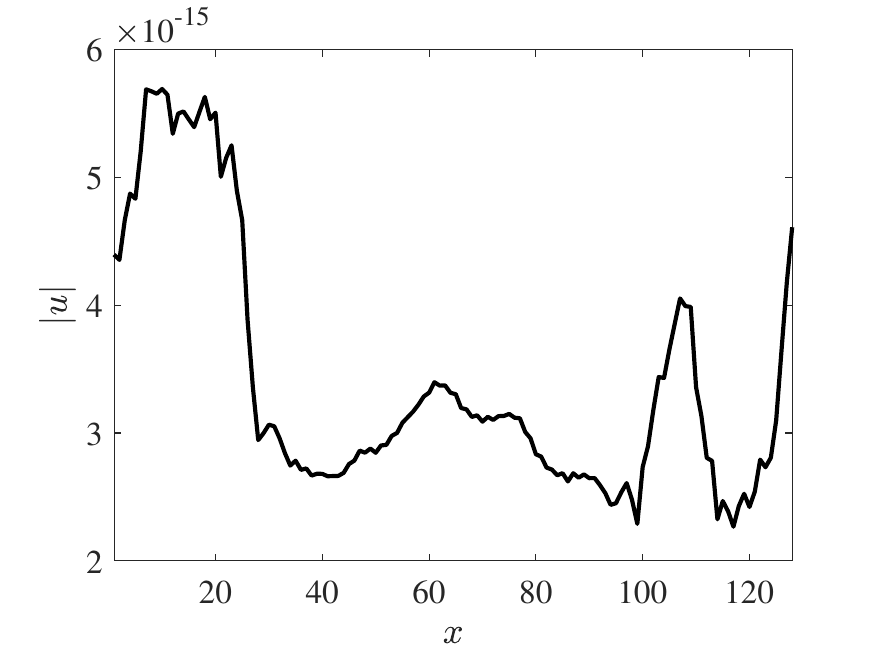}}~~
     \caption{Comparison results of the stationary droplet problem between Yang et al.' model and present model, $\nu_l/\nu_v=1$ and $\rho_l/\rho_v=10$,(a) The time history of the total kinetic energy, (b) chemical potential along $y=0.5L_y$ with the lowest total kinetic energy, (c) velocity profile along $y=0.5L_y$ obtained by the LB model of Yang et al.~\cite{yang2016lattice} with the lowest total kinetic energy, (d) steady velocity profile along $y=0.5L_y$ obtained by the present model.}
     \label{circlecompare}
\end{figure}
The comparison between the present model and Yang's model~\cite{yang2016lattice} is first made.~\cref{circlecompare} shows the time histories of the total kinetic energy, chemical potential distributions and the spurious velocity profiles. Similar to the flat interface problem, the time history of $E$ predicted by Yang's model~\cite{yang2016lattice} exhibits a rapid increase after a period of decrease, which indicates the failure of this model to achieve the well-balanced state, and the minimum velocity from this model is on the order of $O(10^{-9})$. However, the total kinetic energy obtained by the present model finally achieve the well-balanced state with a magnitude of $O(10^{-25})$. In addition, it can be observed that the Yang's model predicts a chemical potential varying at the order of $10^{-3}$, and a velocity field of order $10^{-9}$. However, the chemical potential predicted by the present model is nearly constant and the SV is also reduced to the order of $O(10^{-15})$, showing the well-balanced property of the present model.

\begin{figure}[ht]
     \centering
    \subfloat[]{\includegraphics[width=0.42\textwidth]{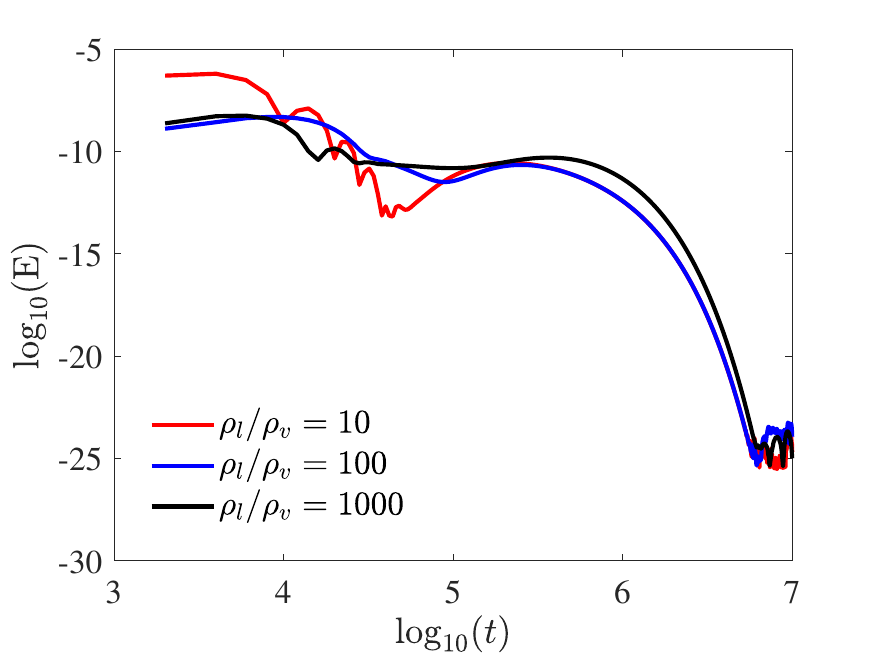}}~~
    \subfloat[]{\includegraphics[width=0.42\textwidth]{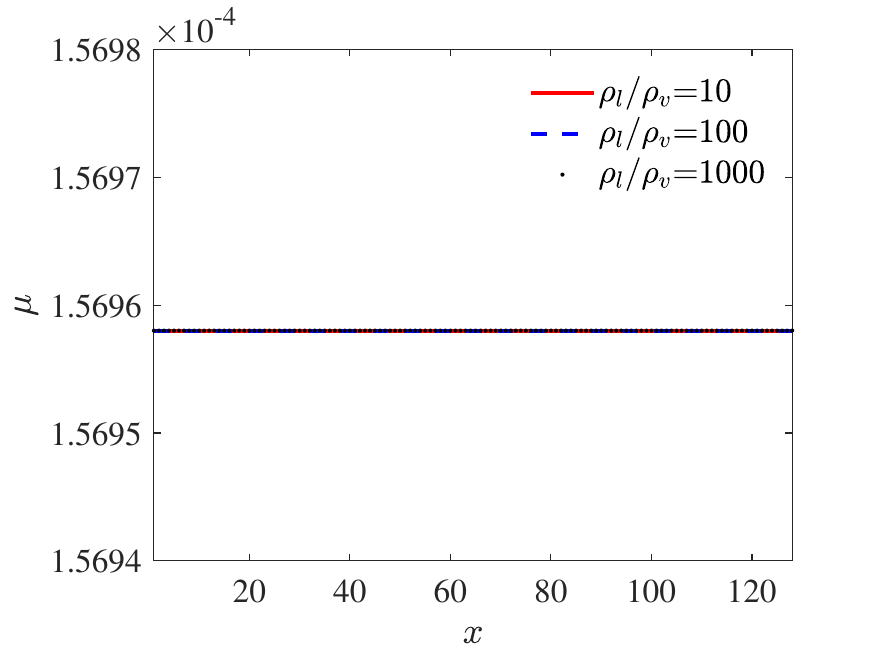}}~~\\
    \subfloat[]{\includegraphics[width=0.42\textwidth]{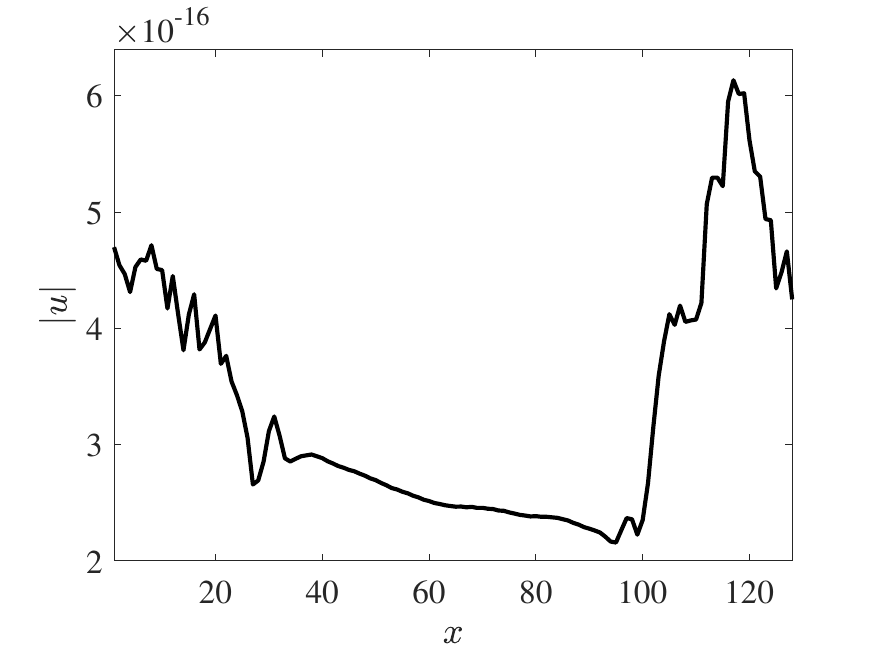}}~~
    \subfloat[]{\includegraphics[width=0.42\textwidth]{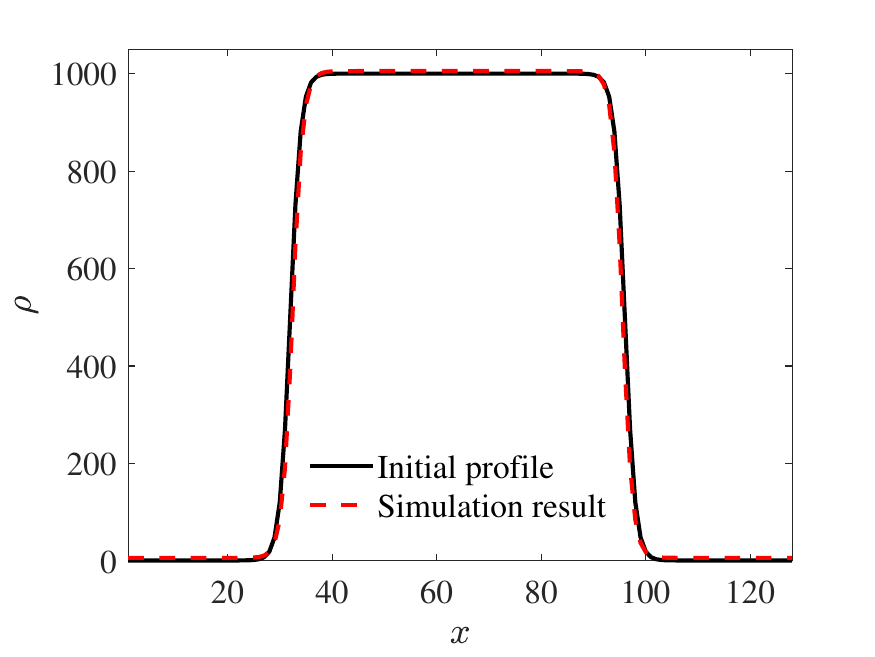}}~~
     \caption{Results of the stationary droplet problem with $\nu_l/\nu_v=1$ and $\rho_l/\rho_v=10,100,1000$,(a) The time history of the total kinetic energy, (b) steady state chemical potential, (c) velocity at steady state with $\nu_l/\nu_v=1$ and $\rho_l/\rho_v=1000$, and (d) density profile along $y=0.5L_y$ with $\nu_l/\nu_v=1$ and $\rho_l/\rho_v=1000$. }
     \label{circlerho}
\end{figure}
Tests with different parameters are further conducted. ~\cref{circlerho}(a) shows the time history of $E$ with different density ratios. It is different from the flat interface problem, the kinetic energies do not increase significantly, but decrease almost monotonically. At the final steady states, the kinetic energy is on the order of $O(10^{-25})$.
~\cref{circlerho}(c) and (d) shows the chemical potential profiles, which suggests that the present model predicted a nearly constant chemical potential.
The velocity profile as well as the density profile along $y=0.5L_y$ with $\rho_l/\rho_v=1000$ and $\nu_l/\nu_g=1$ are also shown in~\cref{circlerho}(c) and (d).
It shows that the density profile from the present model is well captured, and the bulk densities of liquid and vapor are slightly larger than the initial values, which is consistent with the WB-LBE model of Guo~\cite{guo2021well}, the maximum velocity is on the order of $O(10^{-16})$, meaning the SV is vanishes to machine accuracy.

\begin{figure}[ht]
     \centering
    \subfloat[]{\includegraphics[width=0.42\textwidth]{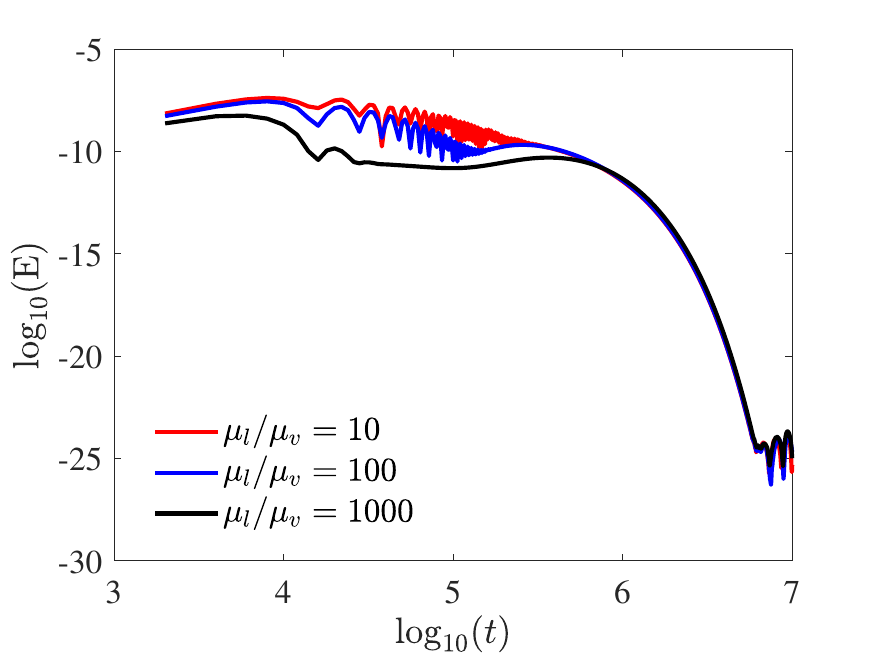}}~~
    \subfloat[]{\includegraphics[width=0.42\textwidth]{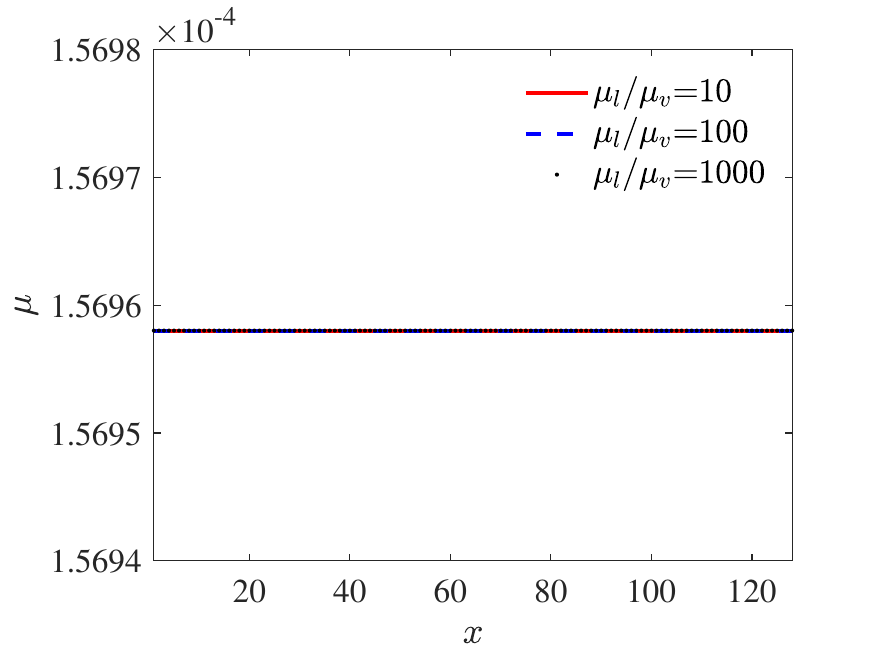}}~~\\
    \subfloat[]{\includegraphics[width=0.42\textwidth]{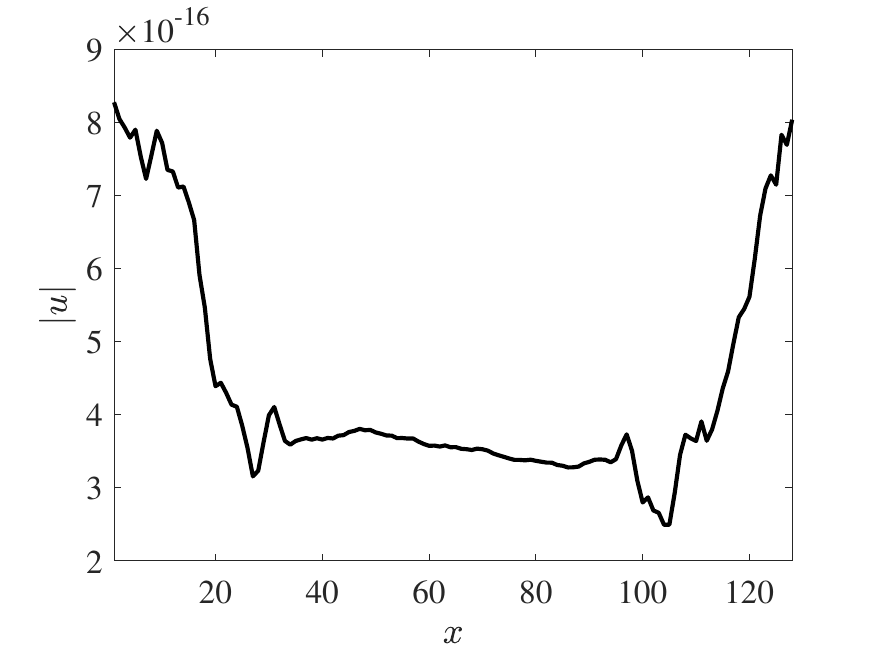}}~~
    \subfloat[]{\includegraphics[width=0.42\textwidth]{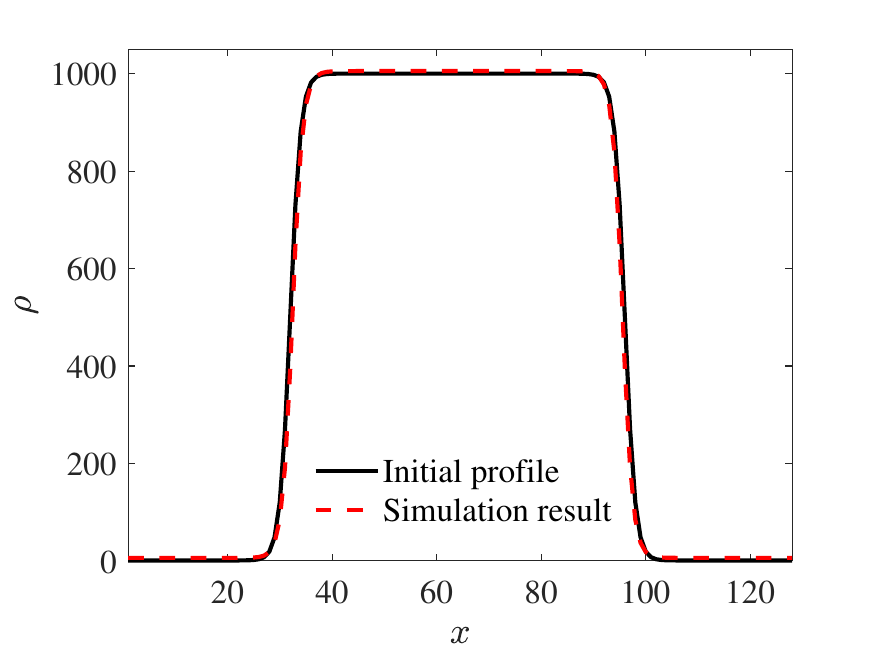}}~~
     \caption{Results of the stationary droplet problem with $\rho_l/\rho_v=1000$ and $\mu_l/\mu_v=10,100,1000$,(a) The time history of the total kinetic energy, (b) steady state chemical potential, (c) velocity at steady state with $\rho_l/\rho_v=1000$ and $\mu_l/\mu_v=100$, and (d) density profile along $y=0.5L_y$ with $\rho_l/\rho_v=1000$ and $\mu_l/\mu_v=100$.}
     \label{circlemu}
\end{figure}
Moreover, a few cases with different viscosity ratios on condition of $\rho_l/\rho_v=1000$ are further considered.
~\cref{circlemu} displayed the predicted results from the present model. It can be seen that both the kinetic energy can be finally reduced to the order of $O(10^{-25})$ in all simulation cases, with the nearly constant chemical poetntial, well fitted density profile and the eliminated SV on the order of $O(10^{-16})$, which demonstrates the good performance of our model for the present two-phase system with a curved interface.

\subsection{The coalescence of two droplets}
The tests of the above cases demonstrate the well-balanced property of the present model for static problems. In this subsection, the coalescence process of two drops is selected to validate the performance of the present model for dynamic problems and demonstrates the improvement on the numerical stability.

All the tests are conducted with the computational domain of size $L_x\times L_y=512\times 512$. The periodic boundary condition is implemented in all directions and the initial configuration consists of two circle droplets, which are positioned in accordance with~\cite{yang2022free}
\begin{equation}
    \phi(x,y)=\frac{\phi_l+\phi_v}{2}+\frac{\phi_l-\phi_v}{2}\left[1-\tanh{\left(\frac{2d_1}{W}\right)}-\tanh{\left(\frac{2d_2}{W}\right)}\right],
\end{equation}
where $d_1$ and $d_2$ are defined as
\begin{equation}
    d_1=\sqrt{(x-x_1)^2+(y-y_1)^2}-R_0, \quad d_2=\sqrt{(x-x_2)^2+(y-y_2)^2}-R_0,
\end{equation}
where $R_0$ donates the radius of droplets, which is set to be $R_0=0.1L_x$. $(x_1,y_1)=(0.5L_x-R_0-0.5W, 0.5L_y)$ and $(x_2,y_2)=(0.5L_x+R_0+0.5W, 0.5L_y)$ are the central positions of the two droplets. Other parameters are set as $M_{\phi}=0.1$, $\alpha=1$, $\sigma=0.005$, $W=8$. The density of the liquid phase is $\rho_l=1$ and the kinematic viscosity of the vapor phase is $\nu_v=0.1$.

\begin{figure}[ht]
     \centering
     {\includegraphics[width=0.7\textwidth]{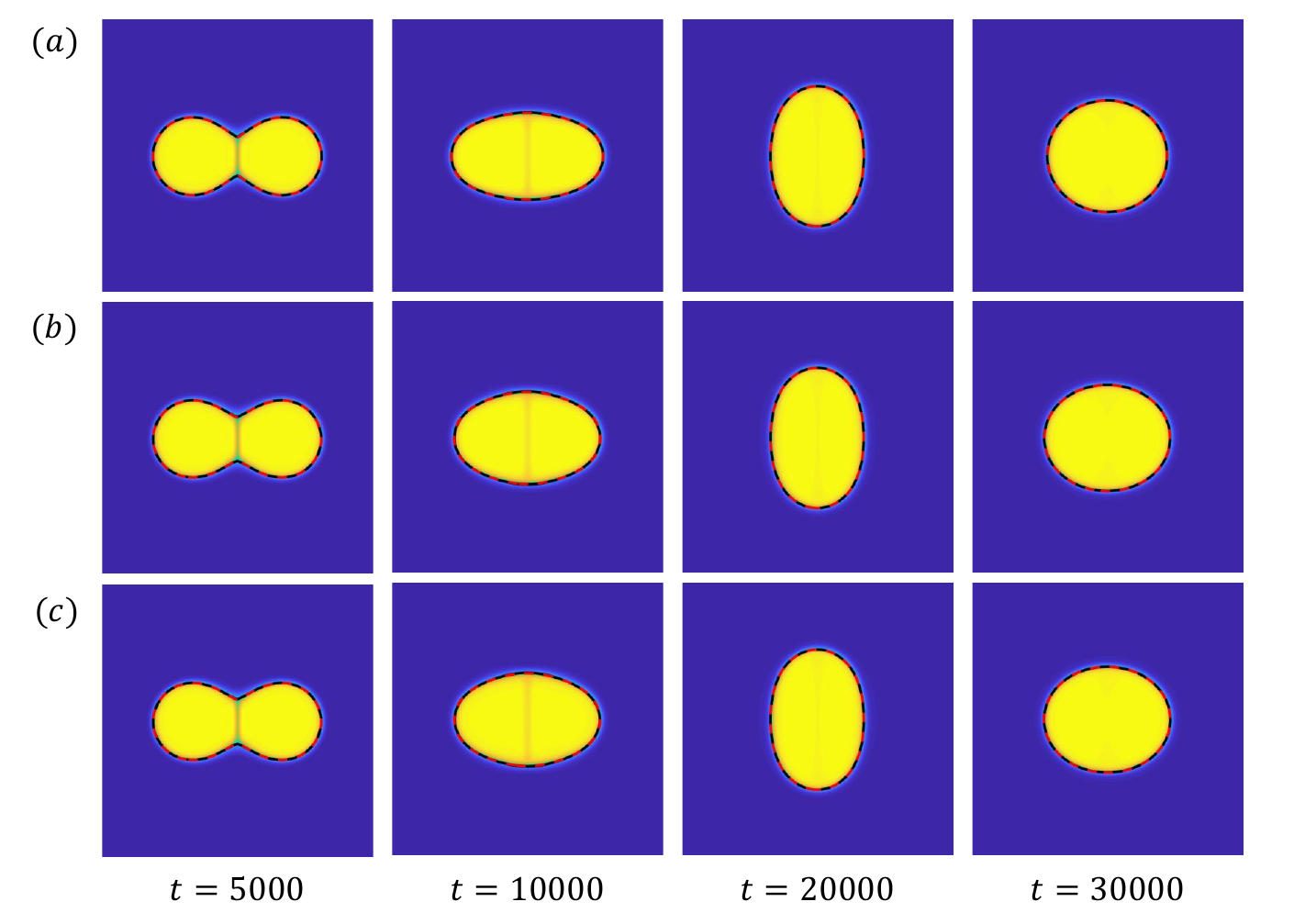}}~~
     \caption{Snap shots for the coalescence of two drops at different time with $\nu_l/\nu_v=1$ and (a) $\rho_l/\rho_v=10$, (b)$\rho_l/\rho_v=100$, (c)$\rho_l/\rho_v=1000$. Diagram of order parameter field: present model; Black solid line: Yang's model~\cite{yang2016lattice} with $\phi=0.5$; Red dash line: Liang's model~\cite{liang2014phase} with $\phi=0.5$.}
     \label{merge1}
\end{figure}
~\cref{merge1} shows the snapshots of the coalescence of two droplets at different times with $\nu_l/\nu_v=1$. Cases with different density ratio i.e., $\rho_l/\rho_v=10,100$ and 1000 are simulated by the present model, Yang's model~\cite{yang2016lattice} and the Liang's model~\cite{liang2014phase} for comparison. It is clearly seen that three models can simulate the coalescence process quite well, and the results from the present model are in good agreement with the reference results, which prove the accuracy of our model for this dynamic problem.

\begin{figure}[ht]
     \centering
     {\includegraphics[width=0.7\textwidth]{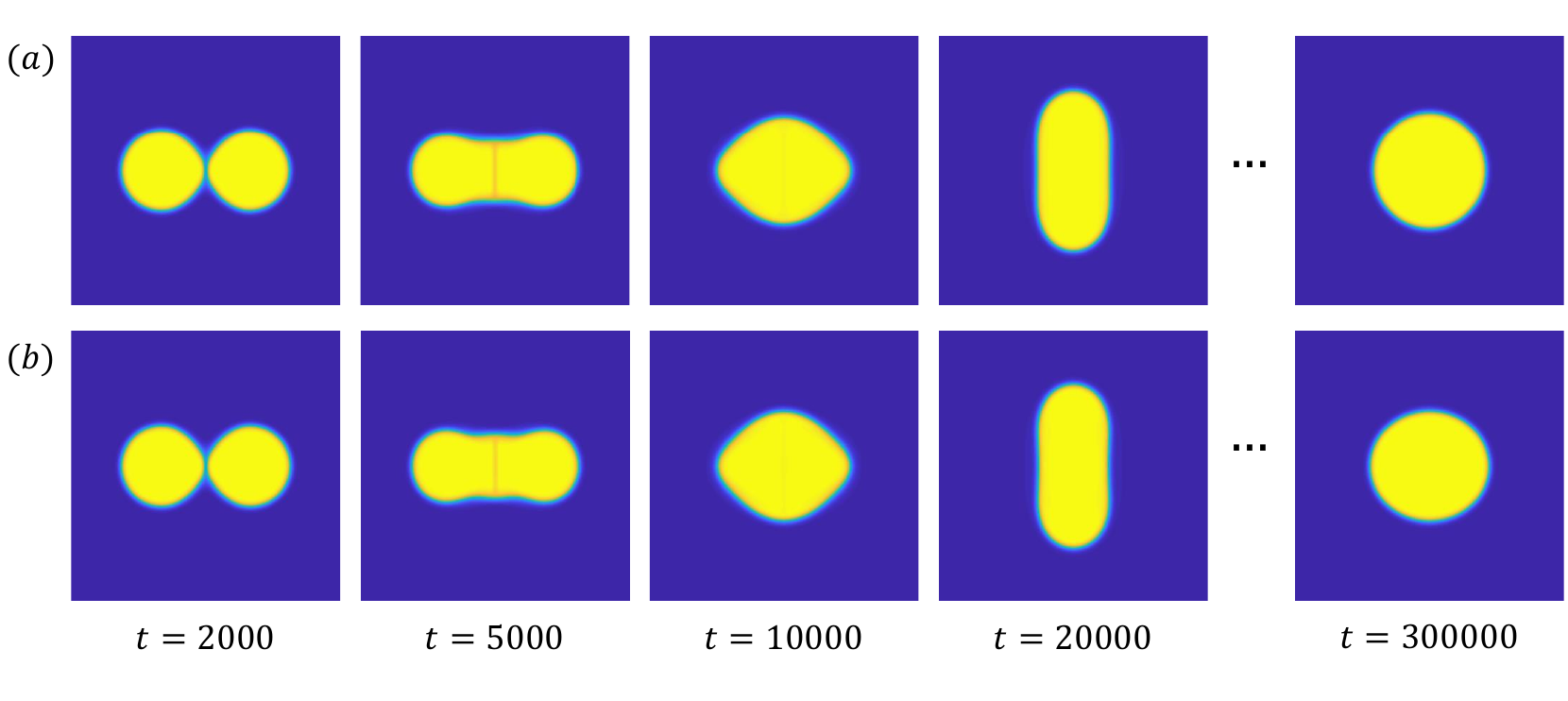}}~~
     \caption{Snap shots for the coalescence of two drops at different time from our present model with $\rho_l/\rho_v=1000$ and (a) $\nu_l/\nu_v=0.1$, (b)$\nu_l/\nu_v=0.01$.}
     \label{merge2}
\end{figure}
Finally we simulate the coalescence processes of two drops with density ratio $\rho_l/\rho_v=1000$ and viscosity ratio $\nu_l/\nu_v=0.1$ and 0.01. Note that the two-phase system with $\rho_l/\rho_v=1000$ and $\nu_l/\nu_v=0.1$ is close to the realistic water-air system at normal atmospheric pressure and room temperature. It is found that both Yang's model~\cite{yang2016lattice} and Liang's model~\cite{liang2014phase} are numerically unstable for this case, but the present model still performs well and the numerical results are displayed in~\cref{merge2}, which demonstrates its good numerical stability.

\section{Conclusion}\label{sec:4}
In this paper, a well-balanced LBE model is proposed for the two-phase fluid system based on the incompressible phase-field theory. The basic idea is to treat the convection term in the CH equation as a separate source term, thereby discarding the velocity term in the equilibrium distribution function. This approach effectively prevents the introduction of additional terms in the recovery to the governing equation through CE analysis. Since the SV arises precisely from the imbalance at the discrete level between the additional terms and the artificial term. This treatment can finally achieve a well-balanced property. Furthermore, such treatment also makes it feasible to attain a divergence-free velocity field to avoid the influence of the artificial compression effects, thus improving the numerical stability. The numerical tests of the flat interface and stationary droplet demonstrate the well-balanced property of the present model, and the test of coalescence of two droplets indicates that two previous similar models~\cite{liang2014phase,yang2016lattice} become numerical unstable for the cases with real physical properties, while the present model still performs well.
This study provides a simple, accurate, and stable LBE model for incompressible two-phase free flow simulations.

\section*{Declaration of Competing Interest}
No Conflict of Interest declared.
\section*{Data Availability Statements}
The data that support the findings of this study are available from the corresponding author upon reasonable request.
\section*{Acknowledgments}
This work was supported by the National Natural Science Foundation of China (No. 51836003 ) and the Interdiciplinary Research Program of Hust (2023JCYJ002). We would like to express appreciation to King Abdullah University of Science and Technology (KAUST) for the support through the grants BAS/1/5028-01 and BAS/1/1423-01-01.

\bibliographystyle{IEEEtr}
\bibliography{ref}
\end{document}